# *CollaboratoR:* A scalable workflow for collaborative data entry and management

Patrick Bills[1,2*], Ashwini Ramesh[1,3,4,5], Laís Petri[1,4,6,7], Alejandra Martínez Blancas[1,4,6,8], Kelly Kapsar[1,3,4], Amar Deep Tiwari[1,9] and Phoebe L. Zarnetske[1,3,4]

*corresponding author: billspat@msu.edu

1.The Institute for Biodiversity, Ecology, Evolution, and Macrosystems (IBEEM), Michigan State University, East Lansing, Michigan, USA 2. The Institute for Cyber-Enabled Research (ICER), Michigan State University, East Lansing, MI, USA 3. Department of Integrative Biology, Michigan State University, East Lansing, MI, 48824, USA 4. Ecology, Evolution and Behavior Program, Michigan State University, East Lansing, MI, 48824, USA 5. Yale Institute of Biospheric Sciences, Yale University, New Haven, CT, USA 6. Department of Plant Biology, Michigan State University, East Lansing, MI, 48824, USA 7. School of Biological Sciences, Southern Illinois University-Carbondale, Carbondale, Illinois, USA 8. Department of Biological Sciences, University of Texas, El Paso, Texas, USA 9. Department of Civil and Environmental Engineering, Michigan State University, East Lansing, MI, 48824, USA

## *Abstract:*

Effective collaborative data entry and transparency are foundational for building robust databases and conducting high-quality data synthesis. Yet, researchers often encounter challenges with inconsistent data entries, inadvertently introducing "smudges" (e.g., errors, misreadings, and inconsistencies) that compromise data integrity. Despite the growing use of open-source tools, many researchers still rely on inefficient formats or costly commercial platforms, whereas fewer adopt complex open-source solutions. These costs and inefficiencies slow down workflows, hindering researchers' ability to efficiently build foundational databases for synthesis research including meta-analyses. To address these challenges, we built upon previous frameworks to develop *collaboratoR*, an open-source, end-to-end workflow that utilizes Google Sheets for collaborative data entry and GitHub for version control, adhering to FAIR data principles. *collaboratoR* fills the gap between ad-hoc spreadsheets and complex systems for data extraction in meta-analyses. We created *collaboratoR* as a customizable R package to automate data validation and aggregation, ensuring consistency and transparency. Data are entered into shared Google Sheets, validated, and pushed to GitHub for version control. After verification, the data are re-validated to ensure accuracy before finalizing. Tested in two case studies—plant competition and avian interaction databases—*collaboratoR* demonstrated its effectiveness in managing large collaborative datasets. Across both case studies, automated validation flagged common entry and formatting issues early, improving traceability and reducing time spent on post-hoc cleaning. This scalable framework can be applied across disciplines where meta-analyses play a critical role in data-driven decision-making, such as social science, ecology, and medical and pharmaceutical research. As a future direction, *collaboratoR's* automated validation can be incorporated into a Continuous Integration (CI) pipeline to run QA/QC on each update and provide rapid feedback to contributors. Ultimately, *collaboratoR* sets a new benchmark for efficient, transparent, and reproducible collaborative data management, minimizing "smudges" and enhancing research synthesis across fields and industries alike.

## Highlights

### *What is already known*

- Collaborative data entry and data transparency are two of the key underpinnings of any synthesis research, including systematic review and meta-analyses.
- Researchers commonly find themselves revisiting and correcting data entries, inadvertently introducing “smudges” (e.g., errors, misreadings, and inconsistencies) that can compromise the integrity of the data.
- Many researchers continue using CSVs, some rely on costly commercial platforms that lack important features for a meta-analysis, and a few others turn to open-source tools that require complex programming, limiting domain experts' control (e.g., non-IT or Computer Science researchers).
- Existing workflows that use GitHub-mediated Quality Assurance/Quality Control (QA/QC) and Continuous Integration (CI) pipelines to automate error checking provide an excellent foundation. However, they are often not end-to-end open source and are rarely implemented to construct databases for meta-analyses.
- These challenges underscore the urgent need for a more accessible and efficient collaborative framework: a low barrier-to-entry and low-cost workflow to streamline collaborative data entry and ensure seamless validation in data management.

### *What is new*

- In the new *collaboratoR* workflow, we designed a pipeline that enables collaborative data entry, tracks changes with the git version control system, and transforms data into “comma separated values” (CSV) files.
- Data are entered in Google Sheets within a collaborative cloud file system, and each file is validated against predefined rules before being pushed to GitHub for tracking.
- This process ensures data consistency and transparency, with all subsequent changes tracked via Git, aligning with FAIR data practices.
- *collaboratoR* incorporates multiple harmonization levels and derived data products, ensuring standardized, reproducible data management across various fields.
- We provide worked examples of how to implement this collaborative workflow for a plant competition meta-analysis and an avian interactions database (AvianMetaNetwork).

### *Potential impact for readers outside the field*

- Discussions of data science invariably describe techniques for cleaning and 'wrangling' existing data sets, but rarely describe the crucial first step of collecting and entering data effectively and accurately in a collaboration.
- The *collaboratoR* workflow is versatile and can be applied across disciplines, using open-source tools like R, CSV formats, and git for end-to-end data management and analysis, while automating the integration of commercial services from Google Workspace and Github for each of use.
- It provides a scalable framework for collaborative data management that can be adopted by various research fields, where data synthesis is pivotal (e.g., social sciences, medicine, and ecology).
- The *collaboratoR* workflow can aid collaborative research efforts in multiple sectors  (e.g., academia, government, non-governmental organizations, and industry); for example, in biomedical companies analyzing data from clinical trials..
- By setting a benchmark for data entry of future synthesis with the *collaboratoR* workflow, we provide a new frontier for research synthesis while also identifying opportunities and new directions for research.

***Introduction**:*

Synthesis research integrates data across studies to address questions at scales no single investigation can reach (Gurevitch et al. 2018). Large-scale collaborative efforts, perhaps most famously the Human Genome Project, have demonstrated the transformative potential of combining data across institutions and disciplines (Collins et al. 2003). Similarly, coordinated research observation networks like the National Ecological Observatory Network (NEON) (Schimel et al 2007) or coordinated experimental networks like Nutrient Network (Borer et al 2014), have effectively scaled up local measurements to continental and global scales (National Academies of Sciences, Engineering, and Medicine, 2025). Realizing this potential, however, depends on a deceptively simple question: how can teams, even with modest group size and budget, collect, enter, and manage shared data reliably? Collaborative data entry harnesses the expertise of diverse teams to handle complex, multifaceted datasets more effectively, especially in fields such as ecology, medicine, and social sciences (Stokols et al 2008; Wuchty et al 2007; Michelle et al 2010; Brun et al 2025). In meta-analysis, for example, researchers with specific knowledge about the studies being reviewed can ensure that data are entered consistently and accurately, despite the nuances of each original study (Gurevitch et al 2018). Working collectively not only speeds up data collection and entry but also provides a mechanism to identify and correct biases in data entry process (Brun et al 2025). Furthermore, collaborative data entry fosters accountability and transparency—principles that are critical in multiple sectors (e.g., academia, government, non-governmental organizations, industry). Several prominent corporations leverage collaborative methods in their meta-analysis services, working with biomedical companies to analyze data across multiple clinical trials (Berlin et al 2019). This approach allows, for example, the comparison of treatment efficacy across multiple clinical trials, ensuring that both the methodological rigor and the relevant domain expertise needed to interpret the results effectively are met (Lane et al 2013). Ultimately, a well-documented collaborative approach can foster accountability, strengthen reproducibility, and accelerate research synthesis.

In many cases, collaborative data entry and subsequent validation are fraught with numerous challenges (Koesten et al 2019). First, version control becomes nearly impossible when data are entered across multiple spreadsheet editors, including Excel and cloud systems. As a result, critical updates are often lost or overlooked (Lakens et al 2016). Lack of version control further exacerbates data integrity issues, especially where access controls or audit trails are absent, leaving data open to errors or tampering (Burchinal & Neebe 2006). Consistent with these risks, over the last two decades, >16,000 (of 49,979) retraction records in scientific literature were attributed to data problems alone (Hu et al 2025). Second, when real-time collaboration tools are lacking, team members often work in isolation, which creates delays, miscommunications, and inconsistencies in data entry (Torka et al 2021; Borer et al 2009). Even with clearly defined protocols, team members may interpret instructions differently, introducing errors that compromise data quality (Winowiecki et al 2011). Third, post-hoc error identification and correction tend to be slow and inefficient. Once the data from multiple studies are merged, errors become more difficult to isolate, often leaving a single person to harmonize the data—a bottleneck that further delays the process (Fowler et al 2017). Fourth, inconsistent formatting and a lack of standardization across contributors create major hurdles for efficient data merging and analysis (Marx 2013). As the size of the dataset increases, it becomes nearly impossible to enter data manually and validate it, with more chances of errors. Finally, compounding these challenges, research indicates that database pre-analysis tasks such as search, retrieval, and development can require an average of 588 hours per study (Allen & Olkin 1999; Marx 2013). This lengthy process creates a significant barrier for researchers trying to resolve data issues during the analysis stage. Taken together, these persistent challenges underscore an urgent need for a new, more streamlined and transparent approach to collaborative data management.

Development platforms already exist to build custom forms and systems for collaborative data entry. However, data entry systems of all types require the deployment of servers to share resources. Even platforms that allow non-programmers to create data entry forms with simple user interfaces require technical expertise to develop and deploy (Yadav 2024). Moreover, many data entry software solutions are designed with commercial use-cases in mind and are often neither open nor web-based, or are platform-specific. As a result, they frequently fall short of meeting the needs of research groups, or require custom programming to provide the features the research groups require, thus forcing all collaborators to work on the same operating system or restrict data entry and management to just a few team members who have purchased licenses. Similarly, open-source tools for building collaborative data entry systems present their own challenges (Trujillo et al 2022). While these systems are flexible and customizable, they often require extensive custom programming, and IT expertise for server deployment and maintenance (Yadav 2024). In our experience, building bespoke collaborative data entry systems in past projects (Mota-Sanchez et al 2002, Turner et al 2018) has revealed the complexities involved in managing these tailored solutions. Relying on custom-built systems requires domain experts to communicate their needs to programmers through a lengthy, iterative development process. This dependency removes control from the researchers, whose core expertise lies in their scientific domain rather than software development, preventing them from directly designing and modifying the data structures they need and ultimately delaying the project and stifling progress (Trujillo et al 2022). Given these challenges, we need an efficient and accessible framework to streamline collaborative data entry and ensure seamless validation across studies.

To tackle these challenges, we developed an end-to-end open source workflow, called ***collaboratoR***, using readily available open and domain-agnostic tools that can be applied across various fields. Our system, built upon a foundation of existing components widely used in science, is an automated data validation and processing workflow to feasibly customize for collaborative data entry without relying on programming or systems administration. For instance, Yenni et al (2019) developed a systematic framework for managing regularly updated ecological data, incorporating GitHub mediated Quality Assurance/Quality Control (QA/QC), and implemented that approach to a near-term dynamic forecasting dataset (White et al 2019). Kim et al (2022) further modified these frameworks by implementing a Continuous Integration (CI) pipeline for both data management and minimizing errors during data collection by field technicians with real-time feedback. This process relies on manually exporting the data which automatically triggers an extensive QA/QC workflow based on automated tools from Github. However, these workflows were designed for primarily field-collected data, require substantial technical infrastructure to deploy, and do not address the collaborative data needs of synthesis research, like meta-analysis. Our *collaboratoR* workflow fills these needs by providing a workflow and automation for processing data in CSV format, while still leveraging easy access Google Sheets and Github APIs for low-cost, user-friendly data entry that researchers can customize without IT expertise. We chose Google Sheets as a service for collaborative data entry for its familiarity, simplicity and availability of automation libraries for the R language From there, data can be automatically exported, validated, and aggregated into CSV files. Importantly, our workflow also preserves multiple tabs during conversion. Prior work has focused on field-collected data (White et al 2019, Kim et al 2022), but we extend this approach to broader data-analysis (e.g, meta-analyses), demonstrating a scalable and flexible workflow relevant to ecology and beyond. By systematizing previously ad-hoc and often error-prone processes, our approach provides a flexible, reproducible framework that makes collaborative data entry and validation more transparent and less cumbersome. In this way, *collaboratoR* complements existing CI-based pipelines, providing a practical stepping-stone for teams that may not have access to specialized infrastructure, while leveraging readily accessible platforms (e.g., Google Sheets APIs) to ensure consistency, traceability, and harmonization across studies.

Here, we introduce and demonstrate the *collaboratoR* workflow by applying it to two case studies: a plant competition database and an avian interaction database. These examples highlight the versatility and effectiveness of the workflow in managing and validating large datasets while ensuring data integrity and transparency. The result is a documented process and software package that is easy to deploy, accessible to researchers with modest technical skills, and flexible enough to be adapted to different project requirements. This scalable framework can be applied across disciplines, from social sciences and ecology to pharmaceuticals, where data-synthesis play a critical role in data-driven decision-making and in the understanding of broad patterns. Ultimately, the goal of the *collaboratoR* is to set a new benchmark for efficient, transparent, and reproducible collaborative data management, minimizing "smudges" and enhancing research synthesis across fields and industries alike.

***Method & Results:*** Below we provide an overview of the *collaboratoR* framework, and describe a workflow that enables collaborative data entry, tracks changes, and transforms data into CSV files.

**1. The Framework**

The *collaboratoR* workflow provides a structured, accessible framework for collaborative data entry, validation, and aggregation across diverse research contexts. The workflow operates through our *collaboratoR* package, integrating multiple functional layers: (1) data retrieval from Google Drive via the *gargle* package (Bryan J, Citro C, Wickham H, 2026); (2) validation using a slightly modified *Validate* package (van der Loo & de Jonge 2021); and (3) conversion of validated entries into CSV files, which are then "committed" via git for reproducible version control. R was chosen for this integration due to its extensive use and familiarity across natural, medical, and social science disciplines. First, for data structure design and entry, we selected Google Sheets, a widely used, researcher-friendly platform, to collaboratively define both the data structure and format requirements. Entry in Google Sheets also allowed us to use "guard rails" around data entry (e.g., enter numeric values in a numeric column) that speeds up the data cleaning. Second, to automate data validation and aggregation, we modified the *validate* package (van der Loo & de Jonge 2021), which facilitates collaborative data entry, tracks changes using git version control, and converts data into CSV files. Data entry occurs in Google Sheets within a shared cloud folder. Each file is validated to ensure it meets predefined rules before being pushed to GitHub for version control. Finally, after data entry, files are pulled, validated again, and pushed to GitHub once verified. This ensures consistency and transparency throughout the data management process. Additionally, the workflow follows the Environmental Data Initiative (EDI) guidelines and adheres to FAIR data principles, incorporating harmonization at multiple levels to ensure standardized, reproducible data management across diverse disciplines.

1.1 EDI framework integration

The *collaboratoR* framework is designed to support a common data structure that aligns with the Environmental Data Initiative (EDI) framework (Gries et al 2023; https://edirepository.org/). This data structure is best understood in three levels, Level 0 (L0): raw data entry, Level (L1): cleaned data, and Level 2 (L2): analysis-ready data. L0 involves iterative refinement when issues arise, L1 focuses on data validation and cleaning, and L2 corresponds to the final analysis and synthesis phase. In accordance, these steps include:

1. Set the study goals to guide data selection and identify the sources.
2. Design the protocol for data collection from the sources based on the study goals and framework (e.g., PRISMA, O'Dea et al 2021).
3. Design the data structures (metadata) required by the protocol, ensuring they align with the goals and data sources.
4. Design the entry system for collecting data based on the design structures.
5. Collect data following the protocol and data entry system. If issues arise with the data source, modify previous steps (L0).
6. Validate the collected data and clean it to a form amenable to analysis (L1).
7. Preliminary analysis of the initial data to confirm the design of goals, collection protocol, and data structures. If necessary, return to steps 3 and 4 and adjust the data structures and entry system.
8. Analysis and summary of collected data (L2).

Our system was created to give domain experts the ability to quickly adjust steps 2-4 with feedback from 5, and to streamline step 6 while providing the added benefit of tracking data provenance, without requiring engineering and deploying complex data technology.

1.2 Version Control of research data with git

Managing changes to data is crucial for teams practicing collaborative, open science. Locally stored spreadsheets like CSVs or Excel sheets or Airtables do not offer built-in tracking features. Although online spreadsheet software provides some version review functionality, this can be lost when files are transferred or copied across different accounts. To overcome these challenges, we use git (https://github.com/git/git), the *de facto* standard for version control, allowing for seamless tracking of changes during collaborative code editing. The git version control system operates on files that are text-based formats, invented to track changes in source code but also applicable to data files. The CSV text-based format is ubiquitous for tabular data and works well with git for version control. If data editors use the standard git 'commit' process as data in a CSV file is added, edited, or deleted, a 'commit message' is permanently attached to those changes in the git database, which is distributed among collaborators and published on a web-based git service such as Gitub. The git visualization of version differences (in command line and as presented on Github) highlights changes made to text files, and changes to CSV files are indicated by row changes. This not only provides transparency for open science, but it also allows one to roll back data files to a previous state if need be.

### 1.3 Why not use a Relational Database Management System?

Relational database management systems (RDBMS) are often used in collaborative research projects to store and manage datasets (for example Yenni et al 2019, Turner et al 2018, Payumo et al 2024, Southworth et al 2026). However the technology requirements for using RDBMS can be a barrier for some users, especially in projects in exploratory stages where data elements are being decided up frequently. An RDBMS requires setup and maintenance of a database and administration of a RDMS injects unneeded complexity that requires technical expertise and time that may not be available in a small team of researchers. In addition, RDBMSs require a custom-made user interface for researchers to efficiently enter data which requires software development and deployment to a shared server or cloud environment which, again, requires software development expertise and time to build these solutions that may not be available in the project team.

The workflow we present here allows for some essential database capabilities such as structured data storage, data integrity (through validation), and relational linkage between tables (such as the list of columns, units of measure, validation rules, support tables like lists of species). However our workflow gives collaborators the ability to quickly adjust the data model by altering the spreadsheet structure and sharing those changes without requiring database administration or custom software development. Frequent low-friction tuning of the data model is crucial in early stages of a data-synthesis.

## 2. The *collaboratoR* workflow

### 2.1 Overview

Having outlined the guiding framework and supporting tools, we next detail the implementation of the *collaboratoR* workflow. The *collaboratoR* workflow, implemented by our *collaboratoR* package, hinges on the configuration and integration of three accessible, free, software platforms: the open-source software R, Cloud (e.g., Google Workspace a.k.a. Google Drive), and version control (GitHub) environments. Based on study goals and protocol, the domain experts first define the necessary data fields, including validation rules and units, and create Data Entry Spreadsheets that adhere to the *collaboratoR* workflows' formatting requirements (described below in Section 2.2). Alongside this, a validation rules file is set up within the R environment to manage more complex rules that match the data definitions. Once the initial setup is complete, domain experts enter preliminary data into the Data Entry Sheets, making adjustments to the fields and rules as necessary. The *validate* package (van der Loo & de Jonge 2021) then reads the Data Entry Schema, validation rules, and Data Entry Sheets to validate the data. If the data passes validation, it is converted to text format and committed to a version control system, such as git. The operator then uploads the latest validated version of this L0 data to a shared repository in GitHub. This workflow effectively bridges the gap between cloud services and version control systems, combining the user-friendly nature of cloud-based spreadsheets with the text-based, code-oriented version control of git, typically CSV data formats.

## 2.2 Workspace configuration

### 2.2.1 Metadata preparation using common spreadsheet software

*Data Entry Schema*: The Data Entry Schema is a Google Sheet that defines the structure of the database. It contains a vertical list of all column names (i.e., fields in the database), along with their data types, descriptions, units, and validation constraints (see Table 1 for required columns). The Data Entry Schema serves as the authoritative reference for what columns are permitted and how they should be formatted. The collaboratoR package reads this sheet and uses it to validate all incoming data. Any changes to column names or data types must be made in the Data Entry Schema first; all other sheets must conform to the structure it specifies.

*Data Entry Sheet*: A Data Entry Sheet is a Google Sheet where collaborators enter raw data. During the study design phase, researchers create these sheets with column names in the first row that match the fields listed in the Data Entry Schema. Column names should be clear and free of special characters that may conflict with downstream software (see Bååth 2012 for best practices for R variable naming conventions). Data Entry Sheets should be separated based on natural groupings amenable to data entry. For example, a multi-species trait database may have one Data Entry Sheet per species, or a meta-analysis may have one Data Entry Sheet per reference paper being analyzed. Within a single Data Entry Sheet, multiple tabs may be used to organize different categories of information (see Case Study I in Section 3 for an illustration). Data Entry Sheets may also use built-in Google Sheets data quality features, such as drop-down menus, to limit entries to predefined options and reduce inconsistencies. For instance, to record geolocation, latitude and longitude columns can be constrained to accept only numeric values in decimal degrees, preventing inconsistencies in format, capitalization, or units. Multiple, related Data Entry Schemas may be used when a project requires different data structures across its Data Entry Sheets.

*Master URL*: The Master URL is a Google Sheet that lists all Data Entry Sheets to be included in the validation process. It must contain at least two columns: one for the URL of each Data Entry Sheet, and one for a unique identifier (ID) for the grouping that the Data Entry Sheet covers. These column names are flexible and are specified when running the R code (e.g., id_column and url_column). The Master URL allows the collaboratoR package to programmatically locate and retrieve all Data Entry Sheets for a given project.

*Dictionary*: A Dictionary is an optional Google Sheet that enumerates the valid entries for a categorical field and stores associated metadata. For example, a project with a "species" column may maintain a species Dictionary listing each species alongside its common name, Latin name, and lifespan. Storing this information in a separate Dictionary rather than repeating it in every Data Entry Sheet reduces redundancy, minimizes the risk of spelling errors, and keeps Data Entry Sheets concise. Dictionaries are created and edited in Google Sheets by domain experts. The collaboratoR package reads them in and uses them during validation to check that categorical entries match the approved set of values.

Together, the Data Entry Schema, Data Entry Sheets, Master URL, and Dictionaries provide the metadata foundation from which the collaboratoR package automatically constructs a validated, uniform database.

| Schema column name | Description | Example from Case study 1 |
|---|---|---|
| col_name | column header which is the field name, following R variable naming rules | "trt_type" |
| col_description | Human readable description of the contents of the column, with reference to the protocol | "competition treatment type, how species are grown: alone, monoculture, mixture" |
| data_str | The data type of the field, conforming to standard types as used by the R 'Validate' package, one of character, integer, numeric, factor | "factor" |
| example | Human readable example data to guide the person entering data entry | "alone, monoculture, mixture" |

**Table 1.** Required columns in the Data Entry Schema read by the *collaboratoR* package

2.2.2 Google Workspace API Configuration (popularly Google Drive)

Google Workspace comprises several tools, including Google docs, sheets, cloud, etc. As it was previously known as Google Drive, and that name remains widely recognized, we will refer to it as Google Drive henceforth. To enable spreadsheet usage, the workflow relies on the Google Sheets package (Bryan J, 2025) which leverages the *gargle* package (Bryan J, Citro C, Wickham H, 2026) to access and download data from Google Drive. If the google sheets are private or protected within an institutional account, this requires a Google Cloud project with the Google Sheets API enabled, following the instructions from the Google Drive developer documentation. Once the Google Cloud account and project are created, one will need to enable APIs and download credential files (*see gargle* documentation for details). However, for those practicing open science and entering data into publicly accessible spreadsheets that anyone can view via a shared link, a separate cloud project becomes unnecessary. This workflow simply requires R code to be read from Google Sheets with read-only access, ensuring accurate and transparent data handling.

2.2.3 Git

The workflow requires the 'git' software to be installed on the computer on which the L0 data will be validated. While many operating systems come with git (e.g., MacOS), collaboration requires a central git server or web-based git service. We used GitHub for this workflow (https://github.com) as it is a free and widely adopted and user-friendly platform. The following steps outline the process: First, make sure that all collaborators who may be using data have a GitHub account (or on the git service of your choice). Second, create a new repository in GitHub to hold CSV data files, which can be either public or private. Ensure that all collaborators who will validate and analyze the data have read and write access. Third, each collaborator should use the standard git process to 'clone' the repository to their working computer in preparation for receiving the validated CSV files. Finally, if the data entry and management are handled separately from the analysis or if multiple analyses are being conducted on the same data set, consider creating two repositories: a primary repository for the entered data (L0 in the EDI framework), and another for pulling these data and managing analysis code. We recommend separating repositories for data validation (L0) and for data wrangling and analysis (L1 and L2) to help maintain a clear division of tasks. This structure allows for modularity and maintainability of the data pipeline, particularly when handling data from multiple sources, though smaller teams may not require this separation.

### 2.2.4 R package/collaboratoR configuration

Besides git software, the R system is required for this workflow. To begin, download our current code (*collaboratoR.Rmd*) from the GitHub repository (URL) as an example to start with, and copy it to a new repository of your own on your computer. While the package website provides detailed instructions for configuration, at a minimum, variables are required to store the Master URLs for the Data Entry Shema (template), the Master URL listing all Data Entry Sheets with their ID value, and one example Google Sheet containing dummy data for testing, and, for convenience, the email address of the user to access Google Drive. If private Google Sheets are used, details from the Google Cloud project, including the Drive API key, must also be stored.

### 2.2.5 Validation Rules

For data validation, this workflow relies on the *validate* package (van der Loo & de Jonge 2021). The *validate* package can read validation rules from a specially formatted file (in YAML format, see package documentation). To create additional validation checks beyond field presence and data type, create a validation file for each type of Data Entry Sheet that follows the Data Entry Schema. Here, we customize the *validate* package for the needs of our datasets. For example, numerical fields can be checked for range or categorical fields to match a controlled vocabulary, and factors can be validated against a list of possibilities (e.g., non-negative biomass, valid country names).

## 2.3 *collaboratoR* workflow

With all necessary setup completed, the workflow is now functional. The following steps describe the process (Figure 1):

1. Select an item or unit of work based on the study requirement. This is the main unit of work, and each unit/item will have its own Google Sheets file. Determine the unique identification value for this unit of work, depending upon the protocol for naming individual Google Sheets. This could be an ID created by the researcher, or a data field such as "species" but must be unique within this project.
2. Create a new, empty Google Sheet for this unit of work by copying the Data Entry Schema template as described above. This Data Entry Sheet must include columns listed in the Data Entry Schema. Name the sheet for the unique ID and add the URL of this Data Entry Sheet to the Master URL, along with the value of the ID.
3. Enter data for this unit following the protocol in the Data Entry Sheet.
4. As the protocol evolves based on issues discovered during data entry, update the Data Entry Schema and the Data Entry Sheets.
5. Periodically, collaborators open the *collaboratoR* package with the configuration for the study, and run the process to update auxiliary metadata if necessary. Then, they run the validation process (e.g., see case studies described below in Section 3) to generate CSVs into the folder containing the L0 data files. This may be done each time a Google Sheet containing data is updated, or for multiple sheets. It is recommended that this process is done regularly.
6. Record the changes to the data using git: the changes to the L0 folder should be saved using the git 'commit' process, adding a message describing what was changed. The collaborator may commit one file at a time, several or all data files at once. Using the commit process for each file changed allows collaborators to write message indicating exactly why the changes were made ('new sheet for group G, or 'fixed typo in value A for ID X')
7. Push these data changes to L0 data in the shared data repository.
8. Collaborators can inspect data changes via the git web service, or by pulling the changes and using the git 'log' and 'diff' functionality.

This workflow ensures data is processed in stages, with clear steps for validation, transformation, and documentation (provenance), leading to harmonized and reusable datasets for research purposes.

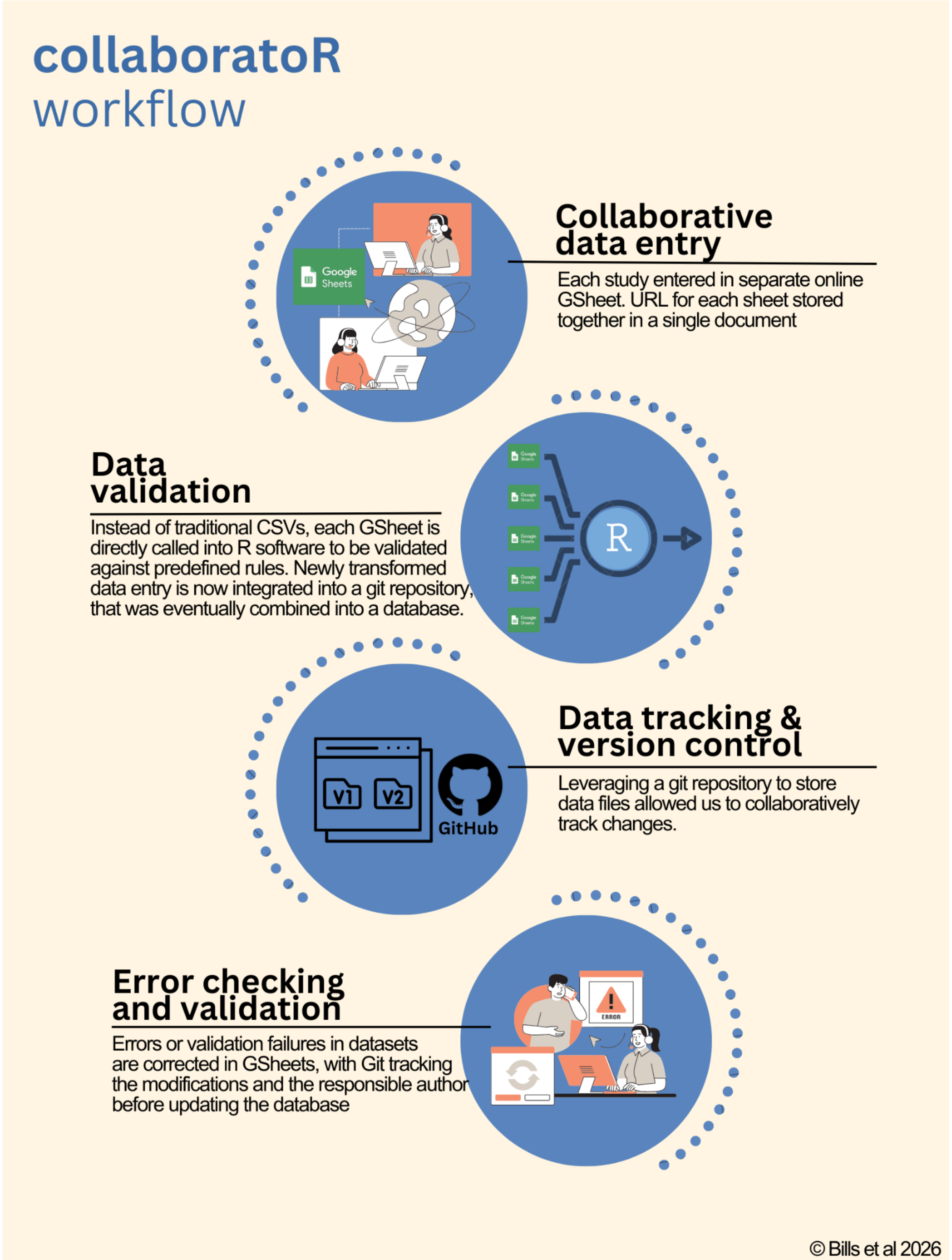

**Figure 1**: The *collaboratoR* workflow, an open-source, end-to-end solution, facilitates collaborative data entry using Google Sheets and version control through GitHub. The *collaboratoR* package automates data validation and aggregation, ensuring data consistency and transparency. Data are entered into shared Google

Sheets, validated, and pushed to GitHub for version control. After verification, the data undergo revalidation to ensure accuracy before finalization. Created in © Canva

**Section 3: Applying the collaboratoR workflow to case studies**

In this section, we highlight the successful implementation of the *collaboratoR* workflow through two collaborative research case studies: a plant competition meta-analysis involving 7 collaborators (Petri, Ramesh et al., 2025) and an avian interaction dataset involving 17 collaborators (Zarnetske et al 2026a, Zarnetske et al 2026b). These examples emphasize the workflow's ability to manage large datasets collaboratively, validate data effectively, and ensure both data integrity and transparency.

3.1 *A plant competition meta-analysis:* Here, we employed the *collaboratoR* workflow to generate a plant competition database which was subsequently used for synthesis and meta-analysis (Ramesh et al 2024). A team of four researchers scanned >3,500 papers on species competition across nutrient levels. Of these, 67 papers met our criteria, with which we constructed a comprehensive database with numerous qualitative and quantitative variables to detail competition outcomes at the species level (Petri, Ramesh et al 2025). Data from each of the 67 studies was entered into Google Sheets, which allowed for organizing different types of information into separate tabs, improving data organization and reducing redundancy. For instance, we isolated biotic data, which contained multiple rows of biomass data from each species, in one tab, while environmental data was stored in a separate tab. This approach helped us avoid repeating environmental data for each new study, resulting in a cleaner, more efficient data structure. Once the data were entered, the links to the files were stored in the Master URL, allowing easy integration with R for subsequent analysis.

For validation, we specifically configured the Google Sheet to enforce a set of rules. For instance, ensuring species and country names were formatted as characters; biomass and variance data were non-negative and numeric; and latitude and longitude values fell within predefined ranges. This early validation process flagged critical missing values, preventing errors from progressing downstream. To maintain oversight, we used version control via GitHub, which allowed us to track changes and identify contributors at each stage. After the bulk of data extraction, we identified the need for collecting information from each species, such as growth forms, life history data, and provenance. However, adding this data to the original Data Entry Sheets at this point would have been both tedious and disruptive, as it would have necessitated changes to the existing columns across multiple Google Sheets. Instead, we developed a separate file, "species dictionary", which could be linked to the species names within the studies, ensuring greater consistency without disrupting the existing structure. After the dataset underwent validation, we exported it to CSV files and combined the individual datasets into a cohesive database. In conventional approaches, dealing with one or two problematic files often led to challenges in identifying the root cause. Here, the validation of each dataset independently made it easy to pinpoint and resolve issues. Only validated files were incorporated into the database, preserving data quality. As a result, we could still write preliminary code and proceed with analysis on the validated data, even if certain datasets failed validation, until everything was fully organized and ready for final processing. By linking these steps, we established a robust framework to construct our database while maintaining data integrity and flexibility, and subsequently used it in a global meta-analysis (Petri, Ramesh et al 2025).

3.2 *An avian interaction dataset:* The AvianMetaNetwork is an open-access dataset consisting of more than 25,000 pairwise bird-bird interactions, for 731 focal bird taxa occurring in the continental United States, Alaska, and Canada during breeding and non-breeding seasons, and 1,258 additional bird taxa that they interact with (Zarnetske et al 2026a, Zarnetske et al 2026b). Biotic interactions, such as competition, predator-prey, brood parasitism, cavity nest creation, mixed flocking, and kleptoparasitism were derived from gray literature and peer-reviewed articles. Information was manually entered into a Google Sheet by data recorders following a strict protocol (Zarnetske et al 2026a). CSV files were downloaded from Google Sheets, entered into GitHub, and verified by a second data recorder prior to entering into the data cleaning workflow. The data cleaning workflow includes further error correction and data verification and harmonization steps to remove typos, update taxonomy, remove duplicates, and ensure intercompatibility of records. Once through the cleaning process, the AvianMetaNetwork consists of an edge list of these interactions, with all records including the citation of the reference and the text excerpt from the reference describing the interaction. This workflow uses Google Drive and git as described above to allow ease of data entry and to track and attribute all data modifications to the user who made them. A major goal of this data collection effort is to align with open science FAIR data principles (Wilkinson et al. 2016).

*Future direction and conclusion:*

Our *collaboratoR* workflow offers numerous opportunities for expansion to improve its functionality and adaptability for a wider range of users and institutions. *First*, our workflow requires all Data Entry Sheets to conform to the Data Entry Schema, but inconsistencies could arise when collaborators alter the Data Entry Sheets inadvertently. A future enhancement could be to implement stricter validation mechanisms that prevent such changes, ensuring the integrity of the data entry process while retaining flexibility. For instance, building on Kim et al. (2022), the workflow could also be extended into a full Continuous Integration (CI) pipeline that automatically runs QA/QC on each update and provides immediate feedback to contributors. *Second*, the use of Google Drive (Cloud) for private (non-open) data entry sheets requires significant setup for many institutions for access by program code. Other enterprise spreadsheets have similar complex requirements (e.g., Microsoft Office). Identifying an alternative data entry system that is nearly as user-friendly and collaborative as commercial spreadsheet services but is easier to use could be a shared community goal. This could involve creating a local file editor that works with CSVs and maintains their structure, and relying strictly on Git/GitHub for data sharing. *Third*, although the validation tool is comprehensive, it is code-oriented. The file format (YAML), while manageable, may not be intuitive for domain experts. One direction for improvement would be to create an easier-to-use syntax for basic validation rules, allowing domain experts to define and manage validation directly within the spreadsheet. *Finally*, the current workflow uses GitHub, and in the future can be expanded to use other service control platforms such as GitLab (https://gitlab.com) or BitBucket (https://bitbucket.org). These services provide additional functionalities such as integrated Continuous Integration and Deployment (CI/CD) workflows, advanced collaboration features, and enhanced project management, offering valuable options for improving the workflow and adapting it to different institutional needs. One caveat is that Git/GitHub can become impractical for very large datasets because tracking many versions inflates repository size; our use case focuses on moderately sized, modular data (e.g., one sheet per study), but other settings may require alternative storage/versioning strategies. Expanding the *collaboratoR* workflow along these lines will not only improve its functionality but also increase its accessibility and efficiency for a broader user base.

*Conclusion*

Successful data synthesis and coordinated field and experiment studies often require collaborative data collection and entry, yet many domain experts lack the engineering and IT skills to effectively use enterprise or custom-made platforms to build a database. Moreover, making interactive changes to data entry protocols typically requires time-consuming modifications to the system, which often involves IT support. This reliance on external expertise limits the ability of domain experts to adapt their systems quickly as they uncover new insights from the data. To address this challenge, the *collaboratoR* workflow provides a scalable and user-friendly solution for collaborative data entry and validation. By leveraging tools like Google Sheets and GitHub, *collaboratoR* empowers domain experts to take full control of their data management, streamlining the process while maintaining data integrity and transparency. Implemented in two case studies—one focused on plant competition and another on an avian database—the workflow has proven effective in managing and validating large datasets that are typically used for a meta-analysis, demonstrating its versatility across disciplines. The impact of *collaboratoR* extends broadly in academic research, offering applications in fields of social sciences, medicine, and ecology, where data synthesis is a critical tool for addressing important research questions. This workflow also holds promise for industries that rely on collaborative methods for data analysis, such as pharmaceuticals and clinical research. By establishing a new benchmark for collaborative data management, *collaboratoR* enhances the efficiency, accuracy, and reproducibility of research synthesis. It opens up new possibilities for researchers to control and adapt their data systems in real-time, ultimately driving more effective and data-driven discoveries across a range of fields.

######################################################################################

***Author Contributions***

Conceptualization: PB (lead), AR, PLZ
Data curation: PB
Investigation: PB
Formal analysis: PB
Methodology: PB
Validation: PB (lead), LP
Visualization: AR
Writing—original draft: PB (equal), AR (equal)
Writing—review and editing: PB, AR, PLZ, AMB, LP, KK, ADT
Supervision: PLZ
Project administration: PLZ

***Competing Interests Authors***: The authors declare that no competing interests exist.

***Code availability:*** https://github.com/IBEEM-MSU/CollaboratoR

***Funding statement:*** This study was funded by The Institute for Biodiversity, Ecology, Evolution, and Macrosystems (IBEEM) (https://ibeem.msu.edu/) at Michigan State University (PB, AR, LP, AMB, ADT, KK, PLZ), through a Michigan State University Strategic Partnership Grant; PB was also supported by Institute for Cyber-Enabled Research (ICER) and a MSU Ecology, Evolution, and Behavior (EEB) Program Seed Grant (to PLZ); AR was also supported by a MSU EEB Presidential Postdoctoral Fellowship.

***Ethics***: The research in this study did not require an ethics review due to the exclusive use of publicly available data and the absence of involvement with human subjects.

***Acknowledgements:*** We thank the Environment Data Initiative for ecological data management vision and guidance, numerous MSU undergraduates instrumental in creating the data and testing the git-based AvianMetaNetwork workflow foundational for this project, the MSU EEB program for supporting this work for providing physical space and financially, and Institute for CyberSupport for supporting PB's involvement in this project. We want to thank the authors of the R "Validate" package for their freely available code and extensive documentation. Google Workspace and Cloud resources and support provided by MSU IT Services Research Cyberinfrastructure.